\documentclass[twocolumn,showpacs,preprintnumbers,amsmath,amssymb]{revtex4}

\usepackage{graphicx}
\usepackage{dcolumn}
\usepackage{bm}
\usepackage{amsmath}
\usepackage{slashbox}

\begin{document}

\def\beq{\begin{equation}}
\def\eeq{\end{equation}}
\def\bea{\begin{eqnarray}}
\def\eea{\end{eqnarray}}

\title{Resummation of Soft Gluon Logarithms in the DGLAP Evolution of Fragmentation Functions}
\author{S. Albino\footnote[5]{Current address: Institut f\"ur Theoretische Physik und Astrophysik,
Universit\"at W\"urzburg, 97074 W\"urzburg, Germany.}}
\affiliation{{II.} Institut f\"ur Theoretische Physik, Universit\"at Hamburg,\\
             Luruper Chaussee 149, 22761 Hamburg, Germany}
\author{B. A. Kniehl}
\affiliation{{II.} Institut f\"ur Theoretische Physik, Universit\"at Hamburg,\\
             Luruper Chaussee 149, 22761 Hamburg, Germany}
\author{G. Kramer}
\affiliation{{II.} Institut f\"ur Theoretische Physik, Universit\"at Hamburg,\\
             Luruper Chaussee 149, 22761 Hamburg, Germany}
\author{W. Ochs}
\affiliation{Max-Planck-Institut f\"ur Physik (Werner-Heisenberg-Institut),\\
F\"ohringer Ring 6, 80805 M\"unchen, Germany}
\date{\today}
\begin{abstract}
We define a general scheme for the evolution of fragmentation functions which resums
both soft gluon logarithms and mass singularities in a consistent manner and to any order, and
requires no additional theoretical assumptions.
Using the Double Logarithmic Approximation and the known perturbative results for the splitting
functions, we present our scheme with the complete contribution from the double logarithms,
being the largest soft gluon logarithms.
We show that 
the resulting approximation is more complete than the Modified Leading Logarithm Approximation
even with the fixed order contribution
calculated to leading order only,
and find, after using it to fit
quark and gluon fragmentation functions to experimental data, 
that this approximation in our scheme gives a good description of the data
from the largest $x_p$ values to the peak region in $\xi=\ln (1/x_p)$, 
in contrast to other approximations.
In addition, we develop a treatment of hadron mass effects
which gives additional improvements at large $\xi$.
\end{abstract}


\maketitle


\section{Introduction}
\label{Intro}

The current description of single hadron inclusive production processes 
within the parton model of perturbative QCD (pQCD)
is provided by fragmentation functions (FFs)
$D_a^h(x,Q^2)$, each of which corresponds at lowest order to the probability
for the parton $a$ produced at short distance $1/Q$ to form a jet that includes 
the hadron $h$ carrying a fraction $x$ of the longitudinal momentum of $a$. 
Different theoretical schemes have been derived depending on the
kinematic region of $x$: Fixed order (FO) calculations at
intermediate and large $x$ and resummation to all orders of soft gluon
logarithms (SGLs) at small $x$. What is needed is a single formalism
valid over the union of all ranges that the different pQCD approaches
allow. This unification must be consistent, i.e. it must agree with each
approach in the set, when the expansion of that approach is used, up to
the order being considered.

Much progress has been made \cite{KKP2000,Albino:2005me} 
in determining fragmentation functions
(FFs) at large and intermediate momentum fraction $x$ 
using FO Dokshitzer-Gribov-Lipatov-Altarelli-Parisi
(DGLAP) evolution \cite{DGLAP} to next-to-leading 
order (NLO) \cite{Curci:1980uw;Furmanski:1980cm}. 
However, determination of FFs at small $x$ is performed independently, since
the calculation of the evolution requires a different
approach because of soft gluon logarithms (SGLs). The complete resummed contribution from the
largest SGLs, being the double logarithms (DLs), of the splitting functions
is obtained from the Double Logarithmic Approximation
(DLA) \cite{Bassetto:1982ma;Fadin:1983aw,Dokshitzer:1991wu}, 
while some information on the contribution from the next largest class of SGLs,
the single logarithms (SLs), in the splitting
functions is obtained from the Modified Leading Logarithm Approximation (MLLA) 
\cite{Dokshitzer:1991wu,Dokshitzer:1984dx,Mueller:1982cq}. (The complete evolution is obtained by using the
approximation that the quark FFs are identical, and equal to the
gluon FF multiplied by a calculable constant.) Provided all
incomplete higher order terms are not allowed to become too large, the
MLLA evolution can describe small $x$ data very well \cite{Albino:2004yg}.
However, although the cross section over a large range of $x$ can be described by pQCD in general, two
different approaches, for which no matching conditions exist, are required
to relate one to the other. In addition,
the range of $x$ over which both approaches are valid is not clear.

In this paper, we are concerned with the unification of DGLAP
evolution and the resummation of SGLs in pQCD. 
In Sec.\ \ref{SGLDGLAP}, we
derive a simple and consistent scheme which reproduces both
approaches to any desired order. 
In Sec.\ \ref{DLADGLAP}, we use the DLA to obtain the complete
DL contribution to the evolution within this scheme. 
These two sections justify and extend the scheme introduced in Ref.\ \cite{Albino:2005gg}.
In Sec.\ \ref{CtD}, we compare this approximation
with experimental data. Finally, in Sec.\ \ref{conclusions}, we present our conclusions.
For the reader's convenience, we list the 
explicit results for the leading order
(LO) splitting functions in the FO approach in
Appendix \ref{app}, as well as 
references for the explicit results for the NLO ones.

\section{SGL Resummation in DGLAP Evolution}
\label{SGLDGLAP}

In this section, we discuss DGLAP evolution and its SGLs in the FO
approach, and then we give a formal definition of our general scheme
in which the SGLs in this approach are resummed.

The DGLAP equation reads
\beq
\frac{d}{d\ln Q^2} D(x,Q^2)=\int_x^1 \frac{dy}{y}P(z,a_s(Q^2)) D\left(\frac{x}{y},Q^2\right),
\label{DGLAPx}
\eeq
where, for brevity, we omit hadron and parton labels. $D$ is a vector
containing the gluon FF $D_g$ and the quark and antiquark FFs $D_q$ and $D_{\overline{q}}$
respectively, in linear combinations according to the choice of basis, and
$P$ is the matrix of the splitting functions.
We define $a_s=\alpha_s/(2\pi)$, whose $Q^2$ dependence is determined
by the QCD $\beta$ function $\beta(a_s(Q^2))$, through
the Callan-Symanzik equation
\beq
\frac{d}{d\ln Q^2}a_s(Q^2)=\beta(a_s(Q^2)).
\label{betafuncdef}
\eeq
The $\beta$ function can be calculated in perturbation theory, where it takes the form
\beq
\beta(a_s)=
-\sum_{n=2}^{\infty}\beta_{n-2} a_s^n.
\label{pseriesforbeta}
\eeq
Choosing a factorization scheme in which $P$ is explicitly independent of quark masses
and then exploiting the resulting SU$(n_f)$ symmetry for $n_f$ quark flavours,
and also exploiting the charge conjugation invariance in $P$, 
leads to the simplest basis for writing Eq.\ (\ref{DGLAPx}), 
which consists of the combinations (i) $D=D_q^-$, where $D_q^-=D_q -D_{\bar{q}}$ is the valence quark FF, 
(ii) $D=D_{NS}$, where $D_{NS}$ is a non-singlet quark FF, i.e.\ any
linear combination of the FFs $D_q^+=D_q+D_{\bar{q}}$ which vanishes when they are all equal,
and (iii) $D=(D_{\Sigma},D_g)$, where 
\beq
D_{\Sigma}=\frac{1}{n_f}\sum_{q=1}^{n_f} D_q^+
\eeq
is the singlet quark FF.
The $n_f-1$ non-singlets must be chosen such that together with the singlet
a linearly independent basis for the quarks is formed.

We will often work in Mellin space, where a function $f(x)$ becomes
\beq
f(\omega)=\int_0^1 dx x^{\omega} f(x),
\label{Meltransdef}
\eeq
where $\omega$ is any integer greater than those values of $\omega$
for which $f(\omega)$ is non-singular,
since the convolution in $x$ space in Eq.\ (\ref{DGLAPx}),
and in equations later in this paper, becomes the simple product
\beq
\frac{d}{d\ln Q^2}D(\omega,Q^2)=P(\omega,a_s(Q^2))D(\omega,Q^2).
\label{DGLAPn}
\eeq
After performing the desired analytic operations, Mellin space results
can be transformed back to $x$ space by analytically continuing $f(\omega)$
to complex $\omega$ and then using the inversion formula
\beq
f(x)=\frac{1}{2\pi i}\int_C d\omega x^{-\omega-1}f(\omega),
\eeq
where $C$ is a contour in Mellin space from ${\rm Im} (\omega)=-\infty$ to ${\rm Im} (\omega)=\infty$, which
passes to the right of all poles in $f(\omega)$. 

Without knowledge of $P$, Eq.\ (\ref{DGLAPx}) (or Eq.\ (\ref{DGLAPn})) provides no
constraint on $D(x,Q^2)$ over the ranges $0\leq x\leq 1$ and $0\leq Q^2 \leq \infty$. 
For a given $\omega$, specification of $P(\omega,Q^2)$ for all $Q^2$ fixes the
the $Q^2$ evolution of $D(\omega,Q^2)$. This means that, given 
$D(\omega,Q_0^2)$ at some specified $Q_0^2$, $D(\omega,Q^2)$ can be calculated for all $Q^2$.
This evolution is usually calculated explicitly for  
the evolution matrix $E$, defined for all $Q^2$ and $Q_0^2$ by
\beq
D(\omega,Q^2)=E\left(\omega,a_s(Q^2),a_s(Q_0^2)\right)D(\omega,Q_0^2),
\label{defofE}
\eeq
and for all $a_s$, $a_1$ and $a_0$ by
\beq
\begin{split}
E(\omega,a_s,a_s)&=I,\\
E(\omega,a_s,a_1)E(\omega,a_1,a_0)&=E(\omega,a_s,a_0),
\label{boundcononE}
\end{split}
\eeq
($I$ is the unit matrix) with no loss of generality for the functional form of $D(\omega,Q^2)$.
Then, with some additional definitions for $E$ that result in no loss of generality,
$E$ becomes fully constrained in terms of $P$ by invoking Eq.\ (\ref{DGLAPn}),
\beq
P(\omega,a_s(Q^2))
=\frac{dE(\omega,a_s(Q^2),a_0)}{d\ln Q^2} E^{-1}(\omega,a_s(Q^2),a_0).
\label{PdefinE}
\eeq
The boundary conditions in Eq.\ (\ref{boundcononE}) can be used to
verify that the right hand side of Eq.\ (\ref{PdefinE}) is independent of $a_0$.

The factorization theorem \cite{Collins:1998rz} states that $P$ is an invariant with respect to the 
hadron being observed, and furthermore that the series for $P(x,a_s)$ in $a_s$ keeping $x$ fixed,
\beq
P(x,a_s)=\sum_{n=1}^{\infty}a_s^n P^{(n-1)}(x),
\label{expanofPinaszfix}
\eeq
can be calculated from perturbation theory even when any quark masses
go to zero or infinity. Equation (\ref{expanofPinaszfix}) truncated at some chosen 
(finite) $n$ is known as the FO approach, and is not valid
at small $x$ due to the presence of terms which
in the limit $x\rightarrow 0$ behave like $(a_s^{n}/x) \ln^{2n-m-1}x$ for 
$m=1,...,2n-1$. Such logarithms are called SGLs, and $m$ labels their class. 
As $x$ decreases, these
unresummed SGLs will spoil the convergence of the FO series for $P(x,a_s)$ once $\ln (1/x) = O(a_s^{-1/2})$.
Consequently the evolution of $D(x,Q^2)$ will not be valid here, since
the whole range $x\leq y\leq 1$ contributes in Eq.\ (\ref{DGLAPx}). Therefore, the FO approach
is only a good approximation for sufficiently large $x$.

SGLs are defined to be all those terms 
of the form $a_s^n/\omega^{2n-m}$ only, where 
$m=1,...,2n$ and labels the class of the SGL, in the expansion about $\omega=0$
of the Mellin transform of Eq.\ (\ref{expanofPinaszfix}),
\beq
P(\omega,a_s)=\sum_{n=1}^{\infty}a_s^n P^{(n-1)}(\omega).
\label{expanofPinasomegafix}
\eeq
We will consider SGLs of the type $m=2n$ later.
For $m=1,...,2n-1$,
this definition agrees with the form of the SGLs in $x$ space given above, since
\beq
\frac{1}{\omega^p}=-\frac{(-1)^p}{p!}\int_0^1 dx x^{\omega}\frac{\ln^{p-1} x}{x}
\label{meltransofsoftglogs}
\eeq
for ${\rm Re} (\omega) >0$ and $p\geq 1$. 
Such terms spoil the convergence of the series in Eq.\ (\ref{expanofPinasomegafix}) 
as $\omega \rightarrow 0$.
What we require is an alternative scheme for the evolution which will be valid for large {\it and}
small $\omega$. The inverse Mellin transform of this evolution should then be valid
for large {\it and} small $x$. For this purpose, we propose the following general scheme, which we 
call the SGL+FO scheme. Firstly, $P$ is written in the form
\beq
P=P^{\rm FO}+P^{\rm SGL},
\label{PsepinSGLsandremain}
\eeq
where $P^{\rm SGL}$ contains only and all the SGLs in $P$,
so that $P^{\rm FO}$ is completely free of SGLs. Secondly, by summing all
SGLs in each class $m$, $P^{\rm SGL}(\omega,a_s)$ is resummed in the form
\beq
P^{\rm SGL}(\omega,a_s)=\sum_{m=1}^{\infty}\left(\frac{a_s}{\omega}\right)^m
g_m \left(\frac{a_s}{\omega^2}\right),
\label{expanpsglinmel}
\eeq
and truncated for some finite $m$. The functions $g_m(x)$ in Eq.\ (\ref{expanpsglinmel}) are 
not Taylor series in either $x$ or any function thereof.
Note that, 
apart from the condition that the series must start at $m=1$, which follows from the
definition of SGLs above,
Eq.\ (\ref{expanpsglinmel}) is just the general result of expanding a function of $a_s$ and $\omega$
in $a_s/\omega$ keeping $a_s/\omega^2$ fixed. 
Thirdly, the remaining FO contribution to $P$,
$P^{\rm FO}(\omega,a_s)$, is expanded in $a_s$ keeping $\omega$ fixed, 
\beq
P^{\rm FO}(\omega,a_s)=\sum_{n=1}^{\infty}a_s^n P^{{\rm FO}(n-1)}(\omega),
\label{expanforPFOinasomega}
\eeq
and truncated for some finite $n$. $P^{\rm FO}(\omega,a_s)$ can be obtained by
subtracting all SGLs from the series for $P$ on the
right hand side (RHS) of Eq.\ (\ref{expanofPinasomegafix}).
Since all classes $m\leq 2n$ are included in $P^{\rm SGL}$, 
i.e. since $P^{\rm SGL}$ contains all terms of the form $a_s^n / \omega^p$ for $p=0,...,2n-1$,
$P^{\rm FO}(\omega,a_s)$ is zero when $\omega=0$, since $P^{{\rm FO}(n)}(0)=0$.
Fourthly and finally, the result for $P(\omega,a_s)$ is inverse Mellin transformed to obtain $P(x,a_s)$, 
and then Eq.\ (\ref{DGLAPx}) is solved exactly (which can be done numerically).

From (incomplete) calculations of Eq.\ (\ref{expanpsglinmel}) up to 
the class $m=2$ \cite{Dokshitzer:1991wu},
$P$ is believed to be finite at $\omega=0$, and in
particular to be a series in $\sqrt{a_s}$ with finite
coefficients, beginning at $O(\sqrt{a_s})$.
This means that each term of the form $a_s^n/\omega^p$ for the types $p\geq 1$
in the expansion of Eq.\ (\ref{expanofPinasomegafix}) about $\omega=0$
should be included in the resummed term of class $m=2n-p$. However, terms of the type
$p=0$ ($m=2n$), which are included in our definition of SGLs, are non-singular and may therefore
be left unresummed. Thus we define a second general scheme, which is the same as the
SGL+FO scheme defined above, except that we separate
$P^{\rm SGL}$ in Eq.\ (\ref{PsepinSGLsandremain}) into
\beq
P^{\rm SGL}=P^{\rm SGL}_{p\geq 1}+P^{\rm SGL}_{p=0},
\label{splitPSGLmeq2}
\eeq 
and expand only $P^{\rm SGL}_{p\geq 1}$ as in Eq.\ (\ref{expanpsglinmel}), while
$P^{\rm SGL}_{p=0}$, which is independent of $\omega$, is expanded as a series in $a_s$.
We shall call this the SGL+FO+FO$\delta$ scheme, where ``+FO$\delta$'' means that the $p=0$
terms, which are each proportional to $\delta (1-x)$ in $x$ space, are left as a FO
series in $a_s$.

To summarize our SGL+FO(+FO$\delta$) scheme, we resum SGLs for which $m=1,...,2n$ ($m=1,...,2n-1$) 
in the form of
Eq.\ (\ref{expanpsglinmel}), and treat all remaining terms as in the FO approach. 

In the phase space region for $a_s \ll 1$ and $x$ above values for which
$\ln (1/x) = O(a_s^{-1/2})$, the following
$x$ space results indicate that the SGL+FO(+FO$\delta$) scheme gives a good description
of the evolution. Firstly,
$P^{\rm SGL}_{p\geq 1}$, obtained from the inverse Mellin transform of Eq.\ (\ref{expanpsglinmel})
without terms of the type $p=0$, can be written as
\beq
P^{\rm SGL}_{p\geq 1}(x,a_s)=\frac{1}{x\ln x}\sum_{m=1}^{\infty}\left(a_s \ln x\right)^m 
f_m \left( a_s \ln^2 x\right).
\label{PSGLinxspaceresummed}
\eeq
Equation (\ref{PSGLinxspaceresummed}) can also be obtained by 
summing the SGLs in $x$ space for each $m$. Since $a_s \ln x$ is always small,
the series in Eq.\ (\ref{PSGLinxspaceresummed}) is a valid approximation 
when $x$ is small.
On the other hand, as $x\rightarrow 1$ the SGLs for the types $p\geq 1$ all vanish, and therefore
so does each term in the series in Eq.\ (\ref{PSGLinxspaceresummed}). Secondly,
the full contribution from the type $p=0$ terms is just
\beq 
P^{\rm SGL}_{p=0}(x,a_s)=\delta(1-x)\sum_{n=1}^{\infty}C_n a_s^n.
\eeq 
Thirdly and finally, the expansion of $P^{\rm FO}(x,a_s)$ in $a_s$,
i.e.\ the inverse Mellin transform of Eq.\ (\ref{expanforPFOinasomega}),
\beq 
P^{\rm FO}(x,a_s)=\sum_{n=1}^{\infty}a_s^n P^{{\rm FO}(n-1)}(x),
\eeq 
converges for all $x$.

\section{DLA Improved DGLAP Evolution}
\label{DLADGLAP}

In this section, we summarize the DLA and its relation to the DGLAP
equation, and use this understanding to derive a modified form of 
DGLAP evolution in which all DLs, the $m=1$ class of SGLs, are included and resummed in a 
manner consistent with the SGL+FO(+FO$\delta$) scheme defined in Sec.\ \ref{SGLDGLAP}.
The DL contribution to the evolution obeys 
the DLA equation, which can be obtained from the DLA master equation for 
the quark and gluon generating functionals given in Ref.\ \cite{Dokshitzer:1991wu},
and is given by
\beq
\begin{split}
\frac{d}{d \ln Q^2}D(x,Q^2)&=\int_x^1 \frac{dy}{y} \frac{2C_A}{y}
a_s(y^2 Q^2) AD\left(\frac{x}{y},y^2 Q^2\right)\\
=\int_x^1 &\frac{dy}{y} \frac{2C_A}{y}A  y^{2\frac{d}{d\ln Q^2}}
\left[a_s(Q^2) D\left(\frac{x}{y},Q^2\right)\right].
\end{split}
\label{DLAx}
\eeq
(The values of various factors are given in Appendix \ref{app}.)
The result $y^{2\frac{d}{d\ln Q^2}}f(Q^2)=f(y^2 Q^2)$ has been used
to obtain the second line in Eq.\ (\ref{DLAx}). Explicitly, $A=0$
for the DL evolving parts of the components $D=D_q^-$ and $D=D_{NS}$, while 
\begin{eqnarray}
A=\left( \begin{array}{cc}
0 & \frac{2 C_F}{C_A} \\
0 & 1
\end{array} \right)
\end{eqnarray}
for the component $D=(D_{\Sigma},D_g)$.
Note that $A$ is a projection operator, i.e.\ it obeys 
\beq
A^2=A.
\label{Aprojection}
\eeq
In Mellin space, Eq.\ (\ref{DLAx}), with certain boundary conditions to be discussed later,
is equivalent to Eq.\ (\ref{DGLAPx}) with $P$ replaced by the $m=1$
term in Eq.\ (\ref{expanpsglinmel}), 
\beq
P^{\rm DL}(\omega,a_s)=\frac{a_s}{\omega}g_1\left(\frac{a_s}{\omega^2}\right),
\label{defofPdlintermsofg1}
\eeq
up to incomplete higher order terms in the remaining FO contribution and up to incomplete
SGLs of classes for which $m\geq 2$.

The remaining part of the evolution (i.e.\ all the FO terms and remaining SGLs)
can be included in Eq.\ (\ref{DLAx}) by writing it in the form
\beq
\begin{split}
\frac{d}{d \ln Q^2}D(x,Q^2)&\\
=\int_x^1 &\frac{dy}{y} \frac{2C_A}{y}A  y^{2\frac{d}{d\ln Q^2}}
\left[a_s(Q^2) D\left(\frac{x}{y},Q^2\right)\right]\\
&+\int_x^1 \frac{dy}{y}\overline{P}(y,a_s(Q^2)) D\left(\frac{x}{y},Q^2\right).
\end{split}
\label{DGLAPandDLA1}
\eeq
$\overline{P}(x,a_s)$, which must be free of DLs, 
is constrained in terms of $P$ since Eq.\ (\ref{DGLAPandDLA1})
must be equivalent to Eq.\ (\ref{DGLAPx})
(for $D=D_q^-$ and $D=D_{NS}$, one obtains the trivial result $\overline{P}=P$). 
In general, this can be done explicitly
by expanding the operator in Eq.\ (\ref{DGLAPandDLA1}) in the form
\beq
y^{2\frac{d}{d\ln Q^2}}=\exp\left[2\ln y \frac{d}{d\ln Q^2}\right]
=\sum_{n=0}^{\infty}\frac{(2\ln y)^n}{n!}\left(\frac{d}{d\ln Q^2}\right)^n
\label{expanofop}
\eeq
and then repeatedly applying the evolution equations, Eqs.\ (\ref{DGLAPx}) and (\ref{betafuncdef}),
to the $\left(\frac{d}{d\ln Q^2}\right)^n \left[a_s(Q^2) D\left(\frac{x}{y},Q^2\right)\right]$ 
operations in Eq.\ (\ref{DGLAPandDLA1}). For example, to $O(a_s^2)$ one finds
\beq
\begin{split}
\overline{P}(x,a_s)=&\ P(x,a_s)-2C_A A \Bigg[\frac{a_s}{x} +2\beta(a_s)\frac{\ln x}{x}\\
&+\int_x^1 \frac{dy}{y}\frac{2a_s x\ln \frac{y}{x}}{y}P(y,a_s)\Bigg]+O(a_s^3).
\end{split}
\label{constraintonPbartoNLO}
\eeq
In the square brackets on the RHS of Eq.\ (\ref{constraintonPbartoNLO}), only the first term
contributes to the $O(a_s)$ (LO) part of $\overline{P}$, while the second and 
third term contribute to the $O(a_s^2)$ part. To this accuracy,
the third term is calculated with
$P(x,a_s)=a_s P^{(0)}(x)$, which can be found in the literature (see Appendix \ref{app}).
From Eq.\ 
(\ref{constraintonPbartoNLO}) we observe that $\overline{P}$ up to $O(a_s^2)$ (NLO)
is free of DLs, by taking the DLs in $P(x,a_s)$ for the
first term on the RHS of Eq.\ (\ref{constraintonPbartoNLO}) up to
$O(a_s^2)$,
\beq
P^{\rm DL}(x,a_s)=2C_A\frac{A}{x}a_s- 4C_A^2 \frac{A\ln^2 x}{x}a_s^2+O(a_s^3).
\label{PDLtoNLOinx}
\eeq

In practice, it is numerically very difficult to solve
Eq.\ (\ref{DGLAPandDLA1}) (and Eq.\ (\ref{DLAx})), in particular 
when $x<Q_0/Q$, which requires calculating the FFs for $Q<Q_0$.
This problem corresponds simply to the fact that in addition to knowing
$D(x,Q_0^2)$, which is required to solve Eq.\ (\ref{DGLAPx}),
$\frac{d}{d\ln Q^2} D(x,Q^2)\big{|}_{Q=Q_0}$ must also be known in order to solve
Eq.\ (\ref{DGLAPandDLA1}). Even more seriously,
Eq.\ (\ref{DGLAPandDLA1}) cannot be solved numerically at all when 
$x\leq \Lambda_{\rm QCD}/Q$, due to the Landau pole in $a_s(y^2 Q^2)$.

Instead, we will examine what constraint Eq.\ (\ref{DGLAPandDLA1})
provides for $P$, since then the evolution can be performed using Eq.\ (\ref{DGLAPx}),
which is numerically easily solved, without
requiring explicit use of $D(x,Q^2)$ at scales less than $Q_0$ at any $x$.
For this purpose, we work in Mellin space, where Eq.\ (\ref{DGLAPandDLA1}) becomes
\beq
\begin{split}
\left(\omega+2\frac{d}{d \ln Q^2} \right)& \frac{d}{d \ln Q^2}D(\omega,Q^2)
=2C_A a_s(Q^2) A D(\omega,Q^2)\\
&\hspace{-0.6cm} +\left(\omega+2\frac{d}{d \ln Q^2}\right)\overline{P}(\omega,a_s(Q^2)) D(\omega,Q^2).
\end{split}
\label{DRAP}
\eeq
After substituting Eq.\ (\ref{DGLAPn}) into Eq.\ (\ref{DRAP}) and dividing out 
the overall factor of  $D(\omega,Q^2)$, we obtain
the following constraint on $P$:
\beq
\left(\omega+2\frac{d}{d\ln Q^2}\right)\left(P-\overline{P}\right)
+2\left(P-\overline{P}\right)P-2C_A a_s A=0.
\label{eqfordelP}
\eeq

Equations (\ref{DGLAPandDLA1}), (\ref{DRAP}) and (\ref{eqfordelP}) 
are exactly equivalent, however Eq.\ (\ref{eqfordelP}) shows most clearly that, for all $Q^2$, 
specifying $\overline{P}(\omega,a_s(Q^2))$ will completely constrain $P(\omega,a_s(Q^2))$ 
once $P(\omega,a_s(Q_0^2))$ is chosen.
However, the only information we have for $\overline{P}$ is that it is free of DLs, 
which means that an explicit constraint can be obtained only for $P^{\rm DL}$, which we now do.
We first make the replacement 
\beq
P=\widetilde{P}+P^{\rm DL}
\label{splitofPintotildePPdl}
\eeq
in Eq.\ (\ref{eqfordelP}), where, comparing with Eq.\ (\ref{PsepinSGLsandremain}), 
\beq
\widetilde{P}=P^{\rm FO}+P^{\rm SGL}-P^{\rm DL}.
\eeq 
Then we expand Eq.\ (\ref{eqfordelP}) as a series in $a_s/\omega$ keeping $a_s/\omega^2$ fixed and 
extract the first, $O((a_s/\omega)^2)$, term to find that
the constraint on $P^{\rm DL}$ is exactly
\beq
2(P^{\rm DL})^2+\omega P^{\rm DL}-2C_A a_s A=0.
\label{DLAeqsimplest}
\eeq
Equation (\ref{DLAeqsimplest}) gives two solutions for each component of $P$. Since
$P$ is never larger than a 2$\times$2 matrix in the basis consisting of
singlet, gluon, non-singlet and valence quark FFs, there are four solutions
which read
\beq
\begin{split}
P^{\rm DL}_{1,\pm}(\omega,a_s)
&=S_{\pm}(\omega,a_s)A,\\
P^{\rm DL}_{2,\pm}(\omega,a_s)
&=-\frac{\omega}{2}I-S_{\pm}(\omega,a_s)A,
\end{split}
\eeq
where
\beq
S_{\pm}(\omega,a_s)=\frac{1}{4}\left(-\omega \pm \sqrt{\omega^2+16C_A a_s}\right).
\eeq
For the evolution matrix $E$ defined in Eq.\ (\ref{defofE}),
these solutions for $P^{\rm DL}$ correspond respectively to
\beq
\hspace{0.38cm}E_{1,\pm}(\omega,a_s,a_0)
=I+\left(e^{R_{\pm}(\omega,a_s,a_0)}-1\right)A,
\label{firstDLAevol}
\eeq
\beq
\begin{split}
E_{2,\pm}&(\omega,a_s,a_0)
=e^{P(\omega,a_s,a_0)}I\\
&-e^{R_{\mp}(\omega,a_s,a_0)}
\int^{a_s}_{a_0}d a \frac{S_{\pm}(\omega,a)}{\beta(a)}
e^{R_{\pm}(\omega,a,a_0)}A,
\end{split}
\label{secondDLAevol}
\eeq
where we have used Eq.\ (\ref{Aprojection}), and defined
\beq
\hspace{0.08cm}R_{\pm}(\omega,a_s,a_0)=\int_{a_0}^{a_s}da \frac{S_{\pm}(\omega,a)}{\beta(a)},
\label{defofR}
\eeq
\beq
P(\omega,a_s,a_0)=-\frac{\omega}{2}\int_{a_0}^{a_s}\frac{da}{\beta(a)}.
\eeq
Here, Eq.\ (\ref{betafuncdef}) has been used to transform the $\ln Q^2$ integrals into integrals
over $a_s$. The general solution to the DL part of 
Eq.\ (\ref{DRAP}) is, finally, Eq.\ (\ref{defofE}) with
\beq
E(\omega,a_s,a_0)=\sum_{i,j}E_{i,j}(\omega,a_s,a_0)k_{i,j}(\omega,a_0),
\eeq
where $i=1,2$ and $j=\pm$.
To ensure that $E$ is normalized as in the first line in Eq.\ (\ref{boundcononE}), 
the matrices $k_{i,\pm}$ obey $\sum_{i,j}k_{i,j}(\omega,a_0)=I$. 
From Eq.\ (\ref{PdefinE}), the most general splitting function is then
\beq
\begin{split}
P^{\rm DL}(\omega,a_s)
=&\left[\sum_{i,j}P^{\rm DL}_{i,j}(\omega,a_s)
E_{i,j}(\omega,a_s,a_0)k_{i,j}(\omega,a_0)\right]\\
&\times\left[\sum_{i,j}
E_{i,j}(\omega,a_s,a_0)k_{i,j}(\omega,a_0)\right]^{-1}.
\end{split}
\label{genformforDLP}
\eeq
The only solution for 
$P^{\rm DL}(\omega,a_s)$ which is consistent with the DLs in the known result for $P$ in the
FO approach of Eq.\ (\ref{expanofPinaszfix}) to $O(a_s^2)$ (see Appendix \ref{app}) is
that for which $k_{1,+}=I$ and all the other $k_{i,j}$ are zero, i.e.\ $P^{\rm DL}=P^{\rm DL}_{1,+}$, or
more explicitly,
\beq
P^{\rm DL}(\omega,a_s)=\frac{A}{4}\left(-\omega+\sqrt{\omega^2+16C_A a_s}\right),
\label{DLresummedinP}
\eeq
since, in the component $D=(D_{\Sigma},D_g)$, 
the expansion of the result in Eq.\ (\ref{DLresummedinP}) in 
$a_s$ to $O(a_s^2)$ keeping $\omega$ fixed gives
\begin{eqnarray}
P^{\rm DL}(\omega,a_s)=
\left( \begin{array}{cc}
0 & a_s \frac{4 C_F}{\omega}-a_s^2 \frac{16 C_F C_A}{\omega^3} \\
0 & a_s \frac{2 C_A}{\omega}-a_s^2 \frac{8 C_A^2}{\omega^3} 
\end{array} \right)+O(a_s^3),
\label{NLODLinmelspace}
\end{eqnarray}
i.e.\ the Mellin transform of Eq.\ (\ref{PDLtoNLOinx}), while in the
components $D=D^-_q$ and $D=D_{NS}$, $P^{\rm DL}=0$.
The other possibilities implied by Eq.\ (\ref{genformforDLP}) 
do not give these results and/or cannot be expanded in $a_s$, i.e. they contain
non-perturbative terms.
Equation (\ref{DLresummedinP}) agrees with the results of Ref.\ \cite{Mueller:1982cq},
which are derived using the conventional renormalization group approach, and with the
results from the generating functional technique of Ref.\ \cite{Dokshitzer:1991wu}. 
Thus, the explicit evolution is given by Eq.\ (\ref{firstDLAevol}) with the upper sign. 
Writing $R_+=R$ and returning to Eq.\ (\ref{defofE}), the evolution
obeys
\beq
D(\omega,Q^2)=\left[I+\left(e^{R(\omega,a_s(Q^2),a_s(Q_0^2))}-1\right)A\right]D(\omega,Q_0^2).
\label{chosenDLAevolsol}
\eeq
For all DLs in the evolution, 
Eq.\ (\ref{chosenDLAevolsol}) solves Eq.\ (\ref{DLAx}).

A completely explicit form for Eq.\ (\ref{chosenDLAevolsol}) can be obtained
in the case that Eq.\ (\ref{pseriesforbeta}) is taken to $O(a_s^2)$ only. 
Equation (\ref{betafuncdef}) then implies that
$a_s(Q^2)=1/(\beta_0 \ln (Q^2/\Lambda_{\rm QCD}^2))$, for which
Eq.\ (\ref{defofR}) (with the upper sign) reads
\beq
\begin{split}
R(\omega,&a_s,a_0)
=\frac{1}{4 \beta_0 a_s}\left(-\omega+\sqrt{\omega^2+16 C_A a_s}\right)\\
&+\frac{2C_A}{\omega \beta_0}\ln 
\frac{\omega+\sqrt{\omega^2+16 C_A a_s}}{-\omega+\sqrt{\omega^2+16 C_A a_s}}
-(a_s \leftrightarrow a_0).
\end{split}
\eeq

Since $P^{\rm DL}_{\Sigma \Sigma}=P^{\rm DL}_{g\Sigma}=0$ according to Eq.\ (\ref{DLresummedinP}), 
it follows from Eq.\ (\ref{DGLAPn}) when $P=P^{\rm DL}$ that, for small $\omega$,
\beq
\frac{d}{d\ln Q^2} D_{\Sigma}(\omega,Q^2)=\frac{P^{\rm DL}_{\Sigma g}(\omega,a_s(Q^2))}
{P^{\rm DL}_{gg}(\omega,a_s(Q^2))} \frac{d}{d\ln Q^2}D_g(\omega,Q^2).
\label{DLArelfordiffDsigmaanddiffDg}
\eeq
With the results for $P^{\rm DL}_{\Sigma g}$ and $P^{\rm DL}_{gg}$ in Eq.\ (\ref{DLresummedinP}),
integrating Eq.\ (\ref{DLArelfordiffDsigmaanddiffDg}) over $\ln Q^2$ gives
\beq
D_{\Sigma}=\frac{2C_F}{C_A}D_g.
\label{DLArelforDsigmaandDg}
\eeq
The constant of integration has been neglected in Eq.\ (\ref{DLArelforDsigmaandDg}), which is valid
for large $Q$.
Equation (\ref{DLArelforDsigmaandDg}) (again up to an additional constant) can also be derived from
Eq.\ (\ref{chosenDLAevolsol}). 

Since the non-singlet and valence quark splitting functions are free of 
DLs, the derivatives of the non-singlet and valence quark FFs with respect to 
$\ln Q^2$ may be neglected at small $\omega$. Again,
integrating such results over $\ln Q^2$ and neglecting the constants of integration
implies that the non-singlet and valence quark FFs vanish. In this case
Eq.\ (\ref{DLArelforDsigmaandDg}) becomes
\beq
D_{q,\overline{q}} =\frac{C_F}{C_A}D_g,
\label{DLArelforDquarkandDg}
\eeq
reducing the number of FFs required for the cross section to one, $D_g$.
Such a low $x$ approximation is often used in DLA or MLLA analyses of data.
However, since we want a complete formalism suitable for both small and large $x$,
we will only use Eq.\ (\ref{DLArelforDquarkandDg}) to partially constrain our choice of parameterization
at low $x$ in the next section.

Complete information on the SL contribution to $P^{\rm SGL}$, given in 
the notation in Eq.\ (\ref{expanpsglinmel}) by
\beq
P^{\rm SL}(\omega,a_s)=\left(\frac{a_s}{\omega}\right)^2
g_2 \left(\frac{a_s}{\omega^2}\right),
\label{defofPSL}
\eeq
cannot be obtained from Eq.\ (\ref{eqfordelP}), since the full SL
contribution to $\overline{P}$ is not known. However, its
SL at $O(a_s)$, which according to
Eq.\ (\ref{constraintonPbartoNLO}) is equal to the SL in $P$ at $O(a_s)$,
given according to Appendix \ref{app} by
\begin{eqnarray}
P^{{\rm SL}(0)}(\omega)=
\left( \begin{array}{cc}
0 & -3C_F \\
\frac{2}{3}T_R n_f \ & -\frac{11}{6}C_A-\frac {2}{3}T_R n_f
\end{array} \right),
\label{singlogsatLO}
\end{eqnarray} 
is a type $p=0$ term (see Sec.\ \ref{SGLDGLAP}). Therefore, in Eq.\ (\ref{eqfordelP}), approximating
$\overline{P}$ by $a_s P^{{\rm SL}(0)}$ should lead to a better approximation for the evolution
than approximating $\overline{P}$ by zero.
Not surprisingly, with this approximation,
Eq.\ (\ref{DRAP}) can be regarded as a generalized version of the MLLA equation to include quarks,
in the sense that the $g$ component of this latter equation for $D=(D_{\Sigma},D_g)$
when Eq.\ (\ref{DLArelforDsigmaandDg}) is invoked is precisely
the MLLA equation of Ref.\ \cite{Dokshitzer:1991wu}. We therefore conclude that 
although Eq.\ (\ref{DRAP}) is derived from the DLA, it is more complete than the
MLLA equation since in Eq.\ (\ref{DRAP}) it is neither 
necessary to restrict $\overline{P}$ in this way nor to use
the approximation in Eq.\ (\ref{DLArelforDsigmaandDg}).

We may now approximately but explicitly calculate the
evolution in the SGL+FO(+FO$\delta$) scheme by
approximating $P^{\rm SGL}$ in Eq.\ (\ref{PsepinSGLsandremain})
by its leading term $P^{\rm DL}$ given by Eq.\ (\ref{DLresummedinP}), being the first,
$O(a_s/\omega)$, term in Eq.\ (\ref{expanpsglinmel}). Thus we take
\beq
P=P^{\rm DL}+P^{\rm FO}(+P^{\rm SGL}_{p=0}).
\label{PsepinDLsandremain}
\eeq
Recall that $P^{\rm FO}$ is equal to $P$ in Eq.\ (\ref{expanofPinasomegafix}) when all SGLs
are excluded. After inverse Mellin transforming Eq.\ (\ref{PsepinDLsandremain}),
we can solve Eq.\ (\ref{DGLAPx}) directly using standard numerical techniques for
$x$ space evolution, as suggested in
Sec.\ \ref{SGLDGLAP}. We shall call this the DL+FO(+FO$\delta$) scheme.
We note that, although
the analytic solution to Eq.\ (\ref{DGLAPn}) for $P=P^{\rm DL}$ 
as defined in Eq.\ (\ref{DLresummedinP}) is given by Eq.\ (\ref{chosenDLAevolsol}),
while the analytic solution to Eq.\ (\ref{DGLAPn}) for $P=P^{\rm FO}$ can be found using
the well known method in the FO approach, an analytic solution to Eq.\ (\ref{DGLAPn})
for $P$ given in Eq.\ (\ref{PsepinDLsandremain}) does not seem feasible.

For the DL+FO(+FO$\delta$) scheme, we require $P^{\rm DL}(x,a_s)$. The
inverse Mellin transform of Eq.\ (\ref{DLresummedinP}) gives
\beq
P^{\rm DL}(x,a_s)=\frac{A\sqrt{C_A a_s}}{x\ln \frac{1}{x}}
J_1\left(4\sqrt{C_A a_s}\ln \frac{1}{x}\right),
\label{allDLinzindelPclosed}
\eeq
where $J_1(y)$ is the Bessel function of the first kind, given by 
\beq
J_1(y)=\frac{y}{2}\sum_{r=0}^{\infty}\frac{\left(\frac{-y^2}{4}\right)^r}
{r! (r+1)!}.
\eeq
This gives
\beq
P^{\rm DL}(x,a_s)=\frac{2C_A a_s A}{x}\sum_{r=0}^{\infty}\frac{(-1)^r}{ r! (r+1)!} (4 C_A a_s \ln^2 x)^r.
\label{allDLinzindelP}
\eeq
The series in Eq.\ (\ref{allDLinzindelP}) may also be obtained by
expanding Eq.\ (\ref{DLresummedinP}) to infinite order in $a_s$, using Eq.\ (\ref{meltransofsoftglogs})
to perform the inverse Mellin transform on each term, and finally using
the identity
\beq
(-4)^r \frac{\left(-\frac{1}{2}\right)\left(-\frac{3}{2}\right)
\cdot \cdot \cdot \left(\frac{3}{2}-r\right)\left(\frac{1}{2}-r\right)}{(2r)!}=\frac{1}{r!}
\eeq
for $r\geq 0$. Equation (\ref{allDLinzindelP}) in fact converges rapidly,
however the truncated series can differ substantially from the 
full series whenever $x$ is small enough. For more reliability,
the Bessel function should be calculated by numerical evaluation of the result
\beq
J_1(y)=\frac{1}{\pi}\int_0^{\pi}d\theta \cos (y\sin \theta -\theta).
\eeq
Note that for $Q>\Lambda_{\rm QCD}$, 
the evolution of $D(x,Q^2)$ in the DL+FO(+FO$\delta$) scheme contains no Landau pole
for all $x$, not just for $x>\Lambda_{\rm QCD}/Q$ as is the case if we evolve with
Eq.\ (\ref{DGLAPandDLA1}).

Using the property at large $y$ that
\beq
J_1(y)=\sqrt{\frac{2}{\pi y}}\cos\left(y-\frac{3\pi}{4}\right) +O\left(\frac{1}{y}\right),
\eeq
we find the true $x\rightarrow 0$ divergence
\beq
\begin{split}
P^{\rm DL}(x,a_s)
&=\frac{(C_A a_s)^{\frac{1}{4}}A}{\sqrt{2\pi} x\ln^{\frac{3}{2}}\frac{1}{x}}
\cos\left(4\sqrt{C_A a_s} \ln \frac{1}{x}-\frac{3\pi}{4}\right)\\
&+O\left(\frac{1}{x\ln^2 \frac{1}{x}}\right)
\end{split}
\eeq
of $P$ to be weaker than the one from the FO approach, which from Eq.\ (\ref{allDLinzindelP})
is proportional to $(1/x)\ln^{2(n-1)}(1/x)$ at $O(a_s^n)$.

For simplicity, in the next section we shall take $P^{\rm FO}$ (and $P^{\rm SGL}_{p=0}$)
to $O(a_s)$ only in the DL+FO(+FO$\delta$) scheme. 
In other words, we will approximate $P$ by
\beq
P=P^{\rm DL}+a_s P^{{\rm FO}(0)}(+a_s P^{{\rm SL}(0)}).
\eeq
We shall call this the DL+LO(+LO$\delta$) scheme. 

It is well known from the MLLA that while DLs give the shape of the 
peak that occurs at small $x$, the gluon component of the type $p=0$ SL is required 
to get the correct peak position. We therefore anticipate that, relative to both the
DL+LO scheme and the MLLA, the DL+LO+LO$\delta$ scheme will give a better description of the 
data since it contains both gluon {\it and} quark components of the type $p=0$ SL.

\begin{figure}[h!]
\includegraphics[width=8.5cm]{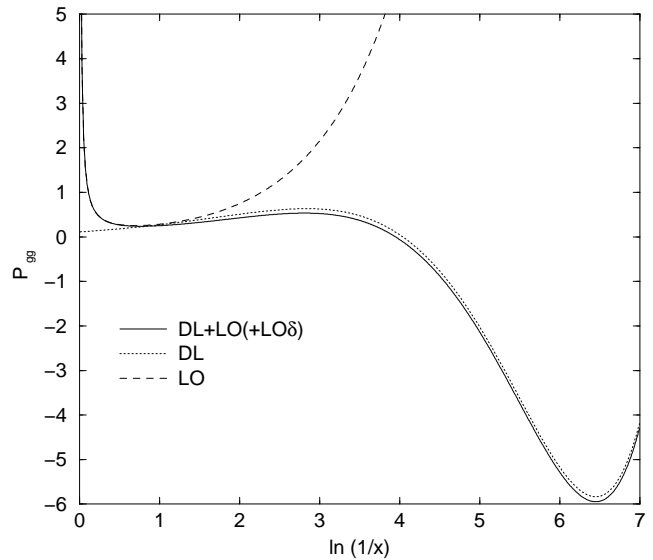}
\caption{\label{plot1} (i) $P_{gg}(x,a_s)$ calculated in the DL+LO(+LO$\delta$) scheme,
(ii) $P_{gg}(x,a_s)$ calculated to $O(a_s)$ in the FO approach (labelled ``LO''), and 
(iii) $P_{gg}^{\rm DL}(x,a_s)$ (labelled ``DL''). $a_s=0.118/(2\pi)$.}
\end{figure}

In Fig.\ \ref{plot1}, we see that $P_{gg}(x,a_s)$  
in the DL+LO scheme, which is equal to the DL+LO+LO$\delta$ scheme when $x\neq 1$, 
interpolates well between its $O(a_s)$ approximation
in the FO approach at large $x$ and 
$P_{gg}^{\rm DL}(x,a_s)$ at small $x$ (the small difference here
comes from $P^{{\rm FO}(0)}(x)$ at small $x$). DL resummation clearly makes a 
large difference to $P$ at small $x$.

\section{Comparisons with data}
\label{CtD}

In this section, we elaborate on
our numerical study of the DL+LO+LO$\delta$ scheme
in Ref.\ \cite{Albino:2005gg}. We perform a numerical comparison
of the FO and DL+LO+LO$\delta$ schemes by otherwise imposing the same 
assumptions and choice of parameterization, starting scale etc., and then fitting
in each scheme to precisely the same experimental measurements
of the normalised differential cross section for light charged hadron production
in the process $e^+ e^- \rightarrow (\gamma,Z) \rightarrow h+X$, where $h$
is the observed hadron and $X$ is anything else. This data, spread over a wide range in
center-of-mass energy $\sqrt{s}$, is composed of the sets from
TASSO at $\sqrt{s}=14$, 35, 44 GeV \cite{Braunschweig:1990yd}
and 22 GeV \cite{Althoff:1983ew}, MARK II \cite{Petersen:1987bq} and
TPC \cite{Aihara:1988su} at 29 GeV, TOPAZ at 58 GeV \cite{Itoh:1994kb}, 
ALEPH \cite{Barate:1996fi}, DELPHI \cite{Abreu:1996na}, L3 \cite{Adeva:1991it},
OPAL \cite{Akrawy:1990ha} and SLC 
\cite{Abrams:1989rz} at 91 GeV, ALEPH \cite{Buskulic:1996tt} and OPAL \cite{Alexander:1996kh} at 133 GeV,
DELPHI at 161 GeV \cite{Ackerstaff:1997kk} and
OPAL at 172, 183, 189 GeV
\cite{Abbiendi:1999sx} and 202 GeV \cite{Abbiendi:2002mj}.
These data span a wide range in
$x_p=2p/\sqrt{s}$, where $p$ is the
momentum of the observed hadron, which constrain the FFs in the region of $x$
for which $x_p\leq x \leq 1$.
In Ref.\ \cite{Albino:2004xa}, large $\xi$ data could be described within the range
\beq
\xi < \ln \frac{\sqrt{s}}{2M},
\label{xicutwithm}
\eeq
where $\xi=\ln (1/x_p)$ and $M$ is a mass scale of $O(1)$ GeV.
In Ref.\ \cite{Albino:2004yg}, this inability to describe
large $\xi$ (small $x_p$) data using the naive approach to the MLLA was attributed to the
formally beyond-MLLA evolutions of the higher moments (skewness, kurtosis etc.) becoming too large at large
$\sqrt{s}$, and when these
were fixed to be zero it was found that the MLLA could give an excellent description of 
data up to the highest values of $\xi$ currently measured. 
In the DL+LO+LO$\delta$ scheme here, the evolution of the higher moments should be somewhat suppressed 
by the FO contribution \cite{Albino:2004yg}. However, we will nevertheless
impose Eq.\ (\ref{xicutwithm}) on the data to be fitted to, since here our aim
is to extend the good FO DGLAP description of small $\xi$ (large $x_p$) data
to larger $\xi$ via DL resummation.

At LO in the coefficient functions, these data are described in terms of the
evolved FFs by
\beq
\frac{1}{\sigma(s)}\frac{d\sigma}{dx_p}(x_p,s)=\frac{1}{n_f \langle Q(s) \rangle }
\sum_q Q_q(s) D_q^+(x_p,Q^2),
\label{approxxs}
\eeq
where $Q_q$ is the electroweak charge of a quark with flavour $q$ and $\langle Q \rangle$ is the average
charge over all flavours. For Eq.\ (\ref{approxxs}) to be a valid approximation,
it is necessary to choose $Q=O(\sqrt{s})$. Since we only use data for which
$\sqrt{s}>m_b$, where $m_b\approx 5$ GeV is the mass of the bottom
quark, and since later we will also set $Q_0>m_b$, we will take $n_f=5$ in all our calculations. 
While the precise choice for $n_f$ does not matter in the DLA, calculations in the FO approach
depend strongly on it. Since we sum over hadron charges, we set $D_{\overline{q}}=D_q$.
Since we do not use data with quark tagging, the $c$ quark cannot be distinguished from
the $u$ quark since both quarks couple to the $Z$ boson in the same way, i.e.\ have the same electroweak
charge. Likewise, the $d$,
$s$ and $b$ quarks are similar to one another in this respect. Therefore,
to avoid redundant degrees of freedom, we fit
only the FFs
\beq
\begin{split}
f_{uc}(x,Q_0^2)&=\frac{1}{2}\left(u(x,Q_0^2)+c(x,Q_0^2)\right),\\
f_{dsb}(x,Q_0^2)&=\frac{1}{3}\left(d(x,Q_0^2)+s(x,Q_0^2)+b(x,Q_0^2)\right)
\end{split}
\eeq
and the gluon $g(x,Q_0^2)$. For each of these three FFs, we choose the parameterization
\beq
f(x,Q_0^2)=N\exp[-c\ln^2 x]x^{\alpha} (1-x)^{\beta},
\label{genparam}
\eeq
since at intermediate and large $x$ the FF is constrained to behave like
\beq
f(x,Q_0^2)\approx Nx^{\alpha} (1-x)^{\beta},
\eeq
which is the standard parameterization used in global fits at large $x$, while at small $x$
(where $(1-x)^{\beta}\approx 1$) the FF is constrained to behave like
\beq
\lim_{x\rightarrow 0}f(x,Q_0^2)=N\exp\left[-c\ln^2\frac{1}{x}-\alpha \ln \frac{1}{x}\right],
\eeq
which for $c>0$ is a Gaussian in $\ln (1/x)$ of width $1/\sqrt{2 c}$,
center at $-\alpha/(2c)$ and normalization given by
$N\sqrt{\pi/c}\exp\left[\alpha^2/(4c)\right]$. For sufficiently large $Q_0$,
the DLA predicts such behaviour with $\alpha <0$. 

In addition, we use Eq.\ (\ref{DLArelforDquarkandDg}) to remove four
free parameters by imposing the constraints
\beq
\begin{split}
&c_{uc}=c_{dsb}=c_g,\\
&\alpha_{uc}=\alpha_{dsb}=\alpha_g.
\label{constraintsonac}
\end{split}
\eeq
This implies that all FFs have the same width and center, however 
the normalisations may not turn out to be consistent with
Eq.\ (\ref{DLArelforDquarkandDg}). However, Eq.\ (\ref{DLArelforDquarkandDg}) is 
only an approximation at small $x$,
while the $N$ are also relevant in the large $x$ region, where
also the $(1-x)^{\beta}$ factors are necessary.
At any rate, it will be interesting to see how well the relation
\beq
\begin{split}
N_{uc} \approx N_{dsb} \approx \frac{C_F}{C_A} N_g
\end{split}
\label{approxrelbetweenNs}
\eeq
as implied by Eq.\ (\ref{DLArelforDquarkandDg}) is obeyed after a fit is performed.
In addition to the 8 free parameters for the FFs, we also fit $\Lambda_{\rm QCD}$.
We choose $Q^2=s$, although it is only important that the latter two quantities
are kept proportional, since the constant of proportionality has no effect on the final
FF parameters and the description of the data (or, equivalently, the quality of the fit). However,
the final fitted $\Lambda_{\rm QCD}$ varies in proportion to this constant, so there will be an overall
theoretical error on our fitted values for $\Lambda_{\rm QCD}$ of a factor of $O(1)$.
Since all data will be at $\sqrt{s}\geq 14$ GeV, we choose $Q_0=14$ GeV. As discussed in
Sec.\ \ref{SGLDGLAP}, the evolution is performed by numerically integrating
Eq.\ (\ref{DGLAPx}). For this we use a grid consisting of 250 points equally spaced  
in $\ln Q^2$ over the range 14 GeV$\leq Q \leq $ 202 GeV, and 
750 points equally spaced in $\ln (1/x)$ over the range $0\leq \ln (1/x) \leq 11.6$. 

\subsection{Fixed Order Evolution}

\begin{figure}[h!]
\includegraphics[width=8.5cm]{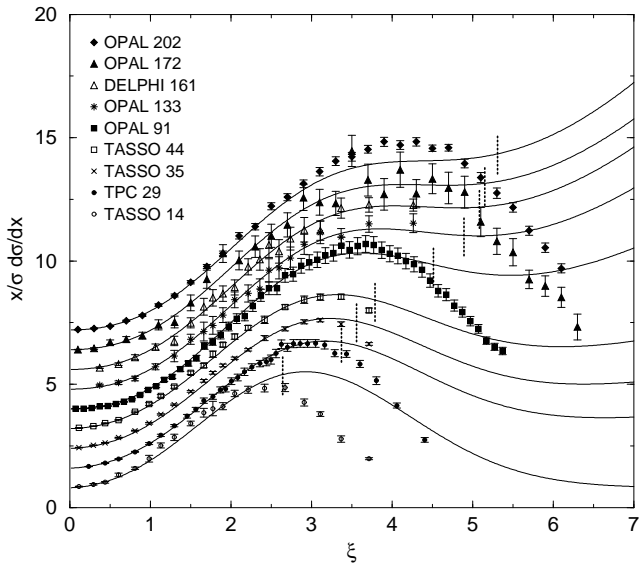}
\caption{\label{fig1} Fit to data as described in Table \ref{tab1}. 
Some of the data sets used for the fit are shown,
together with their theoretical predictions from the results of the fit. Data to the right
of the horizontal dotted lines have not been used in the fit. Each curve is shifted up by 0.8 for clarity.}
\end{figure}

We first perform a fit to all data sets listed above using DGLAP evolution in the FO approach to LO, 
without DL resummation. This approach is the same as that used in fits in the literature.
We fit to those data for which Eq.\ (\ref{xicutwithm}) is obeyed with $M=0.5$ GeV. 
This gives a total of 425 data points out of the available 492. We obtain
a $\chi^2$ per degree of freedom
$\chi^2_{\rm DF}=3.0$ (or 2.1 after subtraction of the contribution to $\chi^2$ from the TOPAZ data,
which is the only data set from which an individual $\chi^2_{\rm DF}$ greater than 6 is obtained), 
and the results are shown in Fig.\ \ref{fig1} and
Table \ref{tab1}. The result for $\Lambda_{\rm QCD}$ is quite consistent with
that of other analyses, at least within the theoretical error.
It is clear that FO DGLAP evolution fails in the description of the peak region and
shows a different trend outside the fit range.
The $\exp[-c\ln^2 x]$ factor
does at least allow for the fit range to be extended to $x$ values below that
of $x=0.1$, the lower limit of most global fits, to around $x=0.05$ ($\xi=3$) for data
at the larger $\sqrt{s}$ values.
Note that the negative value of $\beta$ for the gluon is unphysical,
because the gluon FF is weakly constrained in our fit
since it couples to the data only through the evolution (see Eq.\ (\ref{approxxs})).
\begin{table}[h!]
\begin{footnotesize}
\renewcommand{\arraystretch}{1.1}
\caption{\label{tab1} Parameter values for the FFs at $Q_0=14$ GeV parameterized as in Eq.\
(\ref{genparam}) from a
fit to all data listed in the text using DGLAP evolution in the FO approach to LO.
$\Lambda_{\rm QCD}=388$ MeV.}
\begin{ruledtabular}
\begin{tabular}{c|llll}
\backslashbox{FF}{Parameter} & $N$  & $\beta$ & $\alpha$ & $c$  \\
\hline
                           $g$ & 0.22 & $-$0.43 & $-$2.38  & 0.25 \\
\hline
                         $u+c$ & 0.49 & 2.30    & [$-$2.38]       & [0.25]   \\
\hline
                       $d+s+b$ & 0.37 & 1.49    & [$-$2.38]       & [0.25]   
\end{tabular}
\end{ruledtabular}
\end{footnotesize}
\end{table}

\subsection{Incorporation of Soft Gluon Resummation}

\begin{figure}[h!]
\includegraphics[width=8.5cm]{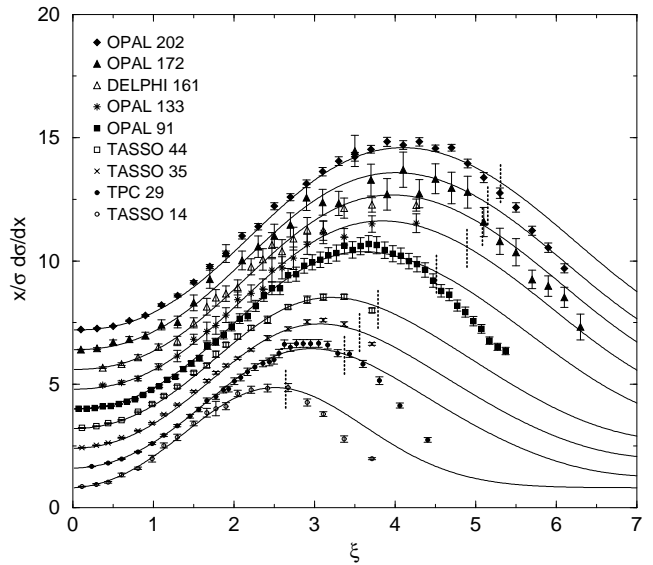}
\caption{\label{fig2} Fit to data as described in Table \ref{tab2}.}
\end{figure}

We now perform the same fit, i.e. to the same data with the same parameterization,
but now evolving in the DL+LO+LO$\delta$ scheme. 
The results are shown in Table \ref{tab2} and Fig.\ \ref{fig2}. 
\begin{table}[h!]
\begin{footnotesize}
\renewcommand{\arraystretch}{1.1}
\caption{\label{tab2} Parameter values for the FFs at $Q_0=14$ GeV parameterized as in Eq.\
(\ref{genparam}) from a fit to all data listed in the text 
using DGLAP evolution in the DL+LO+LO$\delta$ scheme.
$\Lambda_{\rm QCD}=801$ MeV.}
\begin{ruledtabular}
\begin{tabular}{c|llll}
\backslashbox{FF}{Parameter} & $N$  & $\beta$ & $\alpha$ & $c$  \\
\hline
                           $g$ & 1.60 & 5.01    & $-$2.63  & 0.35 \\
\hline
                         $u+c$ & 0.39 & 1.46    & [$-$2.63]       & [0.35]   \\
\hline
                       $d+s+b$ & 0.34 & 1.49    & [$-$2.63]       & [0.35]   
\end{tabular}
\end{ruledtabular}
\end{footnotesize}
\end{table}
We obtain $\chi^2_{\rm DF}=2.1$ (or 1.4 without the TOPAZ data --- for each remaining data set,
the individual $\chi^2_{\rm DF}$ is less than 3), a significant improvement to the fit
above with FO DGLAP evolution. 
The data around the peak is now much better described.
The energy dependence is well reproduced up to the largest $\sqrt{s}$ value, $\sqrt{s}=202$ GeV.
At $\sqrt{s}=14$ GeV, the low $x$ description of the data is extended
from $x=0.1$ in the unresummed case down to 0.06 in the resummed case,
and from $x=0.05$ to 0.005 at $\sqrt{s}=202$ GeV.
This should also be compared 
to the fit to the same data in Ref.\ \cite{Albino:2004xa} where DL resummation
was used within the MLLA but with neither FO terms nor quark freedom (i.e.\ Eq.\ 
(\ref{DLArelforDquarkandDg}) was imposed over the
whole $x$ range). That fit gave $\chi^2_{\rm DF}=4.0$.
We conclude that, relative to the MLLA, the FO
contributions in the evolution, together with freedom from the constraint of Eq.\ 
(\ref{DLArelforDquarkandDg}), makes a significant improvement to the description of the data
for $\xi$ from zero to just beyond the peak.

The value $\Lambda_{\rm QCD}\approx 800$ MeV is somewhat larger than the 
value 480 MeV which we obtain from a DGLAP fit in the large $x$ range ($x>0.1$).
We note that had we made the usual DLA (MLLA) choice
$Q=\sqrt{s}/2$ instead of our choice $Q=\sqrt{s}$ which is usually employed in analyses
using the DGLAP equation, we would have obtained half this value for
$\Lambda_{\rm QCD}$. $N_g$ is too large by a factor of about 2 relative to its prediction
in Eq.\ (\ref{approxrelbetweenNs}). As noted before, the initial gluon FF
is weakly constrained in our fits since it only enters the cross section
via the mixing with the quark FFs in the evolution.

While the DL resummation greatly
improves the description around the peak, the description still deteriorates as $\xi$ increases,
since a fit with data at larger $\xi$ resulted in
an increase in $\Lambda_{\rm QCD}$ and $N_g$, as well as
$\chi^2_{\rm DF}$. The large $\xi$ data may be
better described with the inclusion of the (unknown) resummed SL contribution. 
Indeed, fitting without the LO $p=0$ term (the LO SL), i.e.\
fitting in the DL+LO scheme, generally gives a larger $\chi^2_{\rm DF}$ compared to
the same fits in the DL+LO+LO$\delta$ scheme, since the data around the
peak region could not be described. 

We repeated the above two fits without the constraints given in Eq.\ (\ref{constraintsonac})
and found no substantial improvement to the fits.

The results of these two fits show that, for the DGLAP evolution to describe the overall features
of the data from small to large $x_p$, resummation of SGLs is necessary.
In addition, our DL resummation scheme proves to be an adequate implementation
of this resummation. However, the description at $\xi$ values beyond the peak
remain inaccessible.
It is clear from the smoothness of the shape around and beyond the peak in Fig.\ \ref{fig2} that this cannot
be attributed to any instability in the evolution of the higher moments.
This instability occurred in the method of Ref.\ \cite{Albino:2004xa},
which was pointed out and remedied in Ref.\ \cite{Albino:2004yg}.
Figure \ref{fig2} in fact shows that the evolution at large $\xi$ follows the
data well. Indeed, a fit in the DL+LO+LO$\delta$ scheme 
to data in the same region as that used in the fits of Ref.\ \cite{Albino:2004yg}, 
\beq
\xi >0.75+0.33\ln(\sqrt{s}),
\label{OPALlowercutonxi}
\eeq
gives $\chi^2_{\rm DF}=7$, which is admittedly unacceptably large,
however an FO fit to the same data gives $\chi^2_{\rm DF}=35$.
Thus there is a problem with simultaneously
describing both the range from $\xi=0$ to the peak region to the range beyond the peak
within the DL+LO+LO$\delta$ scheme,
implying a large theoretical error beyond the peak. 
This may be due to the neglect
of the complete SL contribution, which is unfortunately unknown.
Another likely reason is our neglect of the effect of the produced hadron's mass.
This effect is important at small $x_p$.
We will therefore study this effect in the next subsection.

\subsection{Incorporation of Hadron Mass Effects}

We will now incorporate hadron mass effects into our calculations, 
using a specific choice of scaling variable. An alternative approach
is given in Ref.\ \cite{Lupia:1997hj} (see also Ref.\ \cite{Azimov:1985by}).
For this purpose it is helpful to work with light cone coordinates, 
in which any 4-vector $V$ is written in the form 
$V=(V^+,V^-,{\mathbf V_T})$ with $V^{\pm}=\frac{1}{\sqrt{2}}(V^0 \pm V^3)$ and
${\mathbf V_T}=(V^1,V^2)$. In the center-of-mass (COM) frame,
the momentum of the electroweak boson takes the form
\beq q=\left(\frac{\sqrt{s}}{\sqrt{2}},\frac{\sqrt{s}}{\sqrt{2}},
{\mathbf 0}\right). \eeq
In the absence of hadron mass, $x_p$ (whose definition $x_p=2p/\sqrt{s}$
applies only in the COM frame) 
is identical to the light cone scaling variable
$\eta=p_h^+/q^+$. However, the definition $x_p=2p/\sqrt{s}$
applies only in the COM frame,
so $\eta$ is a more convenient scaling variable for studying
hadron mass effects since it is invariant with respect
to boosts along the direction of the hadron's spatial
momentum. Taking this direction to be the 3-axis,
and introducing a mass $m_h$ for the hadron,
the momentum of the hadron in the COM frame reads
\beq p_h=\left(\frac{\eta \sqrt{s}}{\sqrt{2}},\frac{m_h^2}{\sqrt{2}\eta \sqrt{s}},
{\mathbf 0}\right). \eeq
Therefore the relation between the two scaling variables
in the presence of hadron mass is
\beq
x_p=\eta \left(1-\frac{m_h^2}{s\eta^2}\right).
\eeq
Note that these two variables are approximately equal
when $m_h \ll x_p \sqrt{s}$, i.e. hadron mass effects cannot be neglected
when $x_p$ (or $\eta$) are too small.

In the leading
twist component of the cross section after factorization,
the hadron $h$ is produced by fragmentation from a real, massless parton of momentum
\beq k=\left(\frac{p_h^+}{y},0,{\mathbf 0}\right).\eeq
The $+$ component of everything other than this parton and of everything 
produced by the parton other than the observed hadron $h$ must be positive,
implying $y \geq \eta$ and $y \leq 1$ respectively.
As a generalization of the massless case, we assume the 
cross section we have been calculating is $(d\sigma/d\eta)(\eta,s)$, i.e.\ 
\beq
\frac{d\sigma}{d\eta}(\eta,s)=\int_{\eta}^1 \frac{dy}{y} \frac{d\sigma}{dy}(y,s,Q^2)
D\left(\frac{\eta}{y},Q^2\right), \eeq
which is related to the measured observable $(d\sigma/dx_p)(x_p,s)$
via
\beq
\frac{d\sigma}{dx_p}(x_p,s)=\frac{1}{1+\frac{m_h^2}{s\eta^2(x_p)}}
\frac{d\sigma}{d\eta}(\eta(x_p),s).
\eeq

\begin{figure}[h!]
\includegraphics[width=8.5cm]{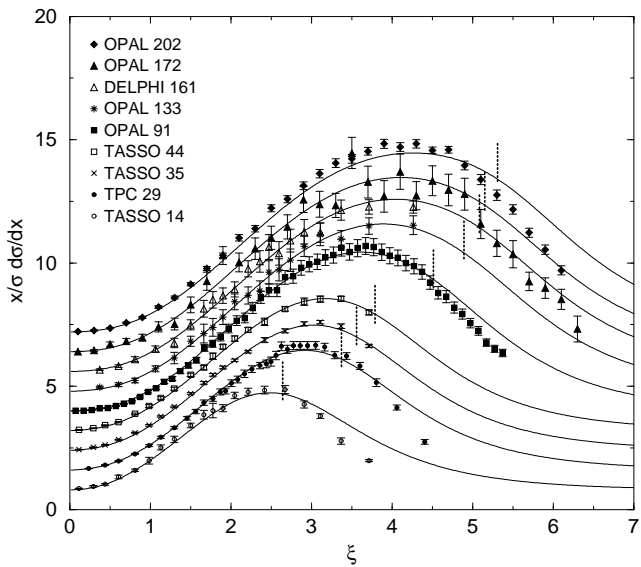}
\caption{\label{fig5} Comparison with data from the fit described in Table \ref{tab3}.}
\end{figure}

Note that the effect of hadron mass is to reduce the size of the cross section
at small $x_p$ (or $\eta$), which Fig.\ \ref{fig2} suggests is what is needed
to improve the fit.
The problem with the above method is that it only applies to the case
where only one species of hadron is produced, whereas the data we are studying 
are for light charged hadrons, i.e.\ charged pions, charged kaons and the 
(anti)protons, whose masses (140, 494 and 938 MeV, respectively)
are substantially different. However, since most of the produced
particles will be pions and kaons, the effective range of hadron masses may be
sufficiently small to achieve reliable results when the 
particle masses are taken to be equal, which we will do.
We now perform the last two fits again but with $m_h$ included in the list of free parameters.
For FO evolution, we obtain $\chi^2_{\rm DF}=2.06$, which is a substantial
improvement over the same fit above for which no treatment of hadron mass effects is applied.
The results are shown in Fig.\ \ref{fig5}
and Table \ref{tab3}. The result for $m_h$ is of the expected order of magnitude,
however $\Lambda_{\rm QCD}$ is unreasonably large. The suppression of the
cross section beyond the peak from hadron mass effects is evident, and allows for the
cross section to follow the data much more closely.
\begin{table}[h!]
\begin{footnotesize}
\renewcommand{\arraystretch}{1.1}
\caption{\label{tab3} As in Table \ref{tab1}, but incorporating mass effects in the fit.
$\Lambda_{\rm QCD}=1308$ MeV and $m_h=408$ MeV.}
\begin{ruledtabular}
\begin{tabular}{c|llll}
\backslashbox{FF}{Parameter} & $N$  & $\beta$ & $\alpha$ & $c$  \\
\hline
                           $g$ & 0.11 & $-$0.82 & $-$2.01  & 0.18 \\
\hline
                         $u+c$ & 0.70 & 2.12    & [$-$2.01]       & [0.18]   \\
\hline
                       $d+s+b$ & 0.82 & 2.35    & [$-$2.01]       & [0.18]   
\end{tabular}
\end{ruledtabular}
\end{footnotesize}
\end{table}

\begin{figure}[h!]
\includegraphics[width=8.5cm]{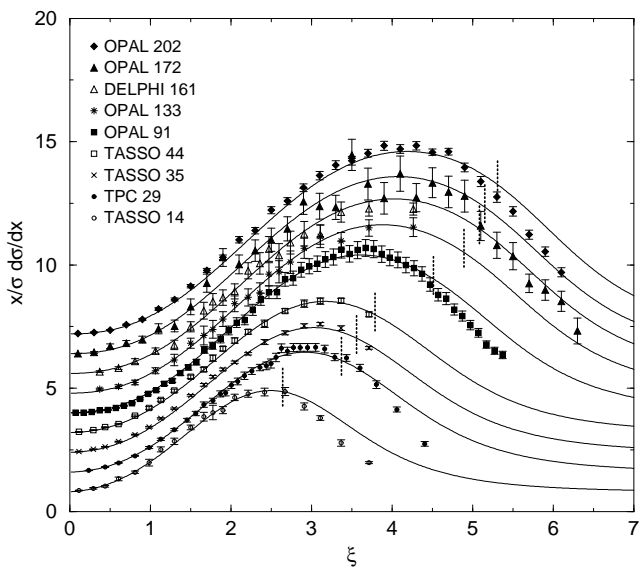}
\caption{\label{fig7} Comparison with data from the fit described in Table \ref{tab4}.}
\end{figure}

For the DL+LO+LO$\delta$ fit, we obtain the results in Fig.\ \ref{fig7} and Table \ref{tab4}.
The parameters are not substantially different to those in Table \ref{tab2}.
The result for $m_h$ is again reasonable. 
We find $\chi^2_{\rm DF}=2.03$, i.e.\ the quality of the fit is the same as
for the previous FO fit, showing that 
the FO case benefits much more from the inclusion of mass
effects than the DL resummed case.
However, treatment of mass effects renders the value of $\Lambda_{\rm QCD}$
obtained in the fit with DL resummation more reasonable.
\begin{table}[h!]
\begin{footnotesize}
\renewcommand{\arraystretch}{1.1}
\caption{\label{tab4} As in Table \ref{tab2}, but incorporating mass effects in the fit.
$\Lambda_{\rm QCD}=399$ MeV and $m_h=252$ MeV.}
\begin{ruledtabular}
\begin{tabular}{c|llll}
\backslashbox{FF}{Parameter} & $N$  & $\beta$ & $\alpha$ & $c$  \\
\hline
                           $g$ & 1.59 & 7.80 & $-$2.65  & 0.33 \\
\hline
                         $u+c$ & 0.62 & 1.43    & [$-$2.65]       & [0.33]   \\
\hline
                       $d+s+b$ & 0.74 & 1.60    & [$-$2.65]       & [0.33]   
\end{tabular}
\end{ruledtabular}
\end{footnotesize}
\end{table}

We conclude therefore that to improve the large $\xi$ description and
to achieve a reasonable value for $\Lambda_{\rm QCD}$, both DL resummation
and treatment of mass effects are required.

To extend NLO calculations to small $x_p$, the complete resummed DL contribution
given by Eq.\ (\ref{allDLinzindelPclosed}) must be added
to the NLO splitting functions. These contain
SGLs belonging to the classes $m=1,...,4$, which must be subtracted.
Note that the NLO $m=1$ term is accounted for by the resummed DL contribution.
The $m=4$ term is a type $p=0$ term, and hence does not need to be subtracted.

\subsection{Comparison With Gluon Jet Data}

\begin{figure}[h!]
\includegraphics[width=8.5cm]{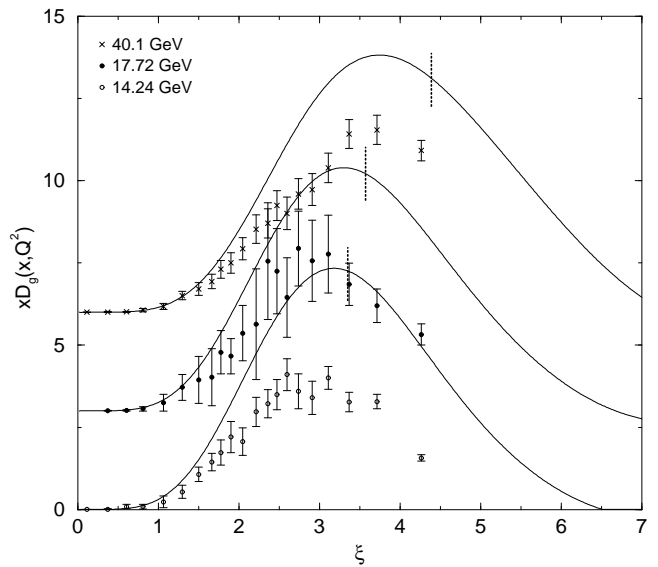}
\caption{\label{fig3} Comparison of the gluon FF 
in the DL+LO+LO$\delta$ scheme to gluon jet measurements from OPAL.
The postion of the cut of Eq.\ (\ref{xicutwithm}) is shown to indicate where
the gluon is constrained. $Q=2E_{\rm jet}$.}
\end{figure}

\begin{figure}[h!]
\includegraphics[width=8.5cm]{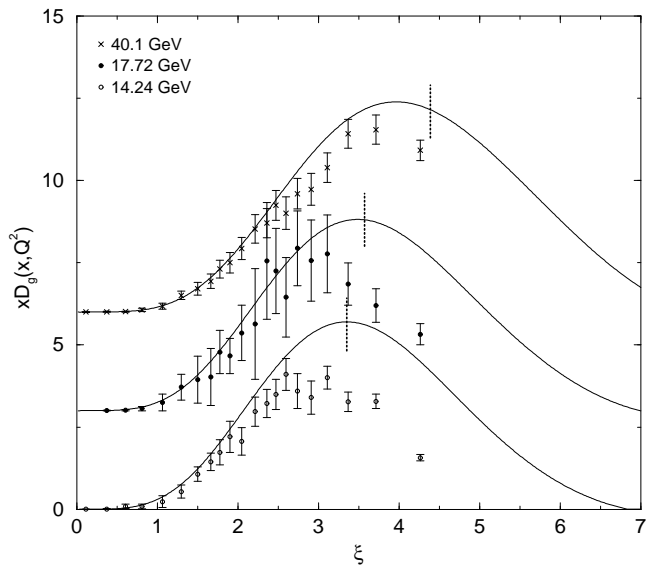}
\caption{\label{fig4} As in Fig.\ \ref{fig3}, but from a fit
in which the gluon jet data below the cut are included in the fit.}
\end{figure}

\begin{figure}[h!]
\includegraphics[width=8.5cm]{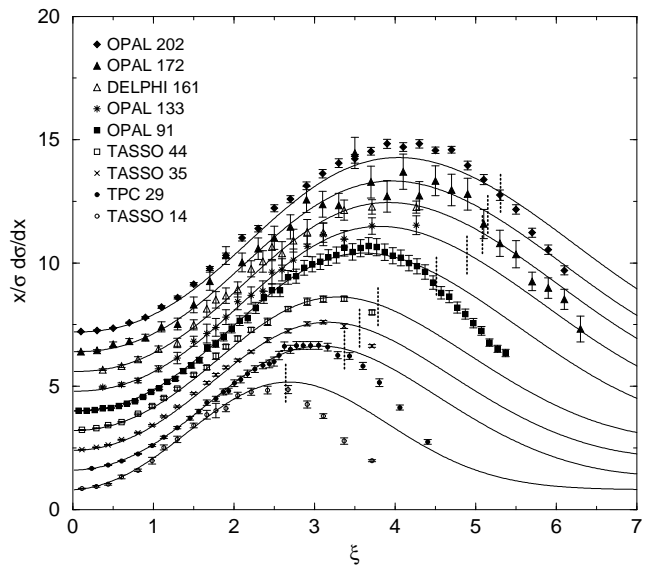}
\caption{\label{fig8} Fit to data as described in Table \ref{tab5}.}
\end{figure}

We now compare our results with the OPAL gluon jet measurements at
$E_{\rm jet}=14.24$, 17.72 \cite{Abbiendi:2003gh} 
and 40.1 GeV \cite{Abbiendi:1999pi}. These data are shown in Fig.\ \ref{fig3},
together with our gluon FF from the fit of Table \ref{tab2} incorporating
DL resummation. The initial gluon FF from that fit
is found to be about twice as large
as predicted by Eq.\ (\ref{approxrelbetweenNs}), and this is reflected in
the figure.
In Fig.\ \ref{fig4}, we show the same plot again, but this time
from a fit in which these data are included (but which is
otherwise identical to the fit of Table \ref{tab2}). The data
is identified with our evolved gluon FF at $Q=2E_{\rm jet}$. In this case
$\chi^2_{\rm DF}=2.3$, a slightly larger value than that from
the fit of Table \ref{tab2}, in particular because 
the gluon jet data at $E_{\rm jet}=14.24$ GeV
cannot not be well fitted. These data give an individual
$\chi^2_{\rm DF}$ of 3.7, although the data at $E_{\rm jet}=17.72$ and 40.1 GeV
give 0.9 and 1.2 respectively. The data around and beyond the peak are poorly
described. The parameters and comparison
with the remaining data are shown in Table \ref{tab5} and Fig.\ \ref{fig8}.
The value for $N_g$ is lower than that in Table \ref{tab2}, and
in better agreement with Eq.\ (\ref{approxrelbetweenNs}).
This smaller gluon FF is presumably the cause of the undershoot
of the calculation from the data at high $\sqrt{s}$ seen in Fig.\ \ref{fig8}.
\begin{table}[h!]
\begin{footnotesize}
\renewcommand{\arraystretch}{1.1}
\caption{\label{tab5} Parameter values for the FFs at $Q_0=14$ GeV parameterized as \
in Eq.\
(\ref{genparam}) from a fit to all data, including the OPAL 
gluon jet data
using DGLAP evolution in the DL+LO+LO$\delta$ scheme.
$\Lambda_{\rm QCD}=954$ MeV.}
\begin{ruledtabular}
\begin{tabular}{c|llll}
\backslashbox{FF}{Parameter} & $N$  & $\beta$ & $\alpha$ & $c$  \\
\hline
                           $g$ & 1.27 & 4.21    & $-$2.41  & 0.29 \\
\hline
                         $u+c$ & 0.40 & 1.55    & [$-$2.41]       & [0.29]   \\
\hline
                       $d+s+b$ & 0.50 & 1.59    & [$-$2.41]       & [0.29]
\end{tabular}
\end{ruledtabular}
\end{footnotesize}
\end{table}

\begin{figure}[h!]
\includegraphics[width=8.5cm]{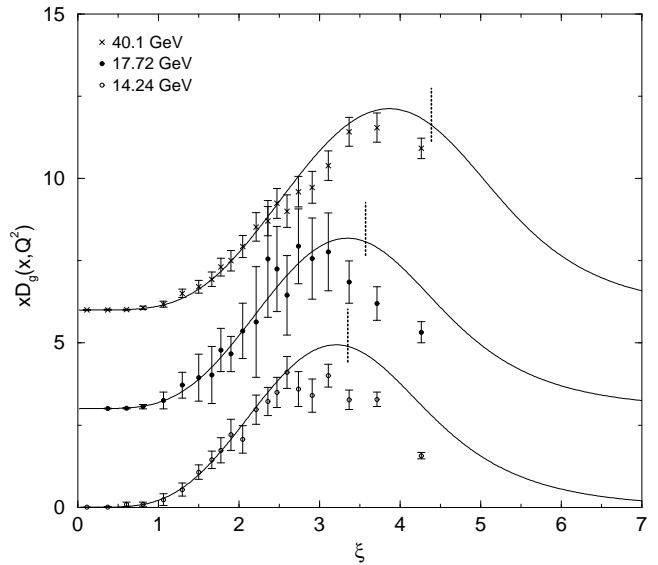}
\caption{\label{fig6} As in Fig.\ \ref{fig4}, but with a fitted hadron mass.}
\end{figure}

\begin{figure}[h!]
\includegraphics[width=8.5cm]{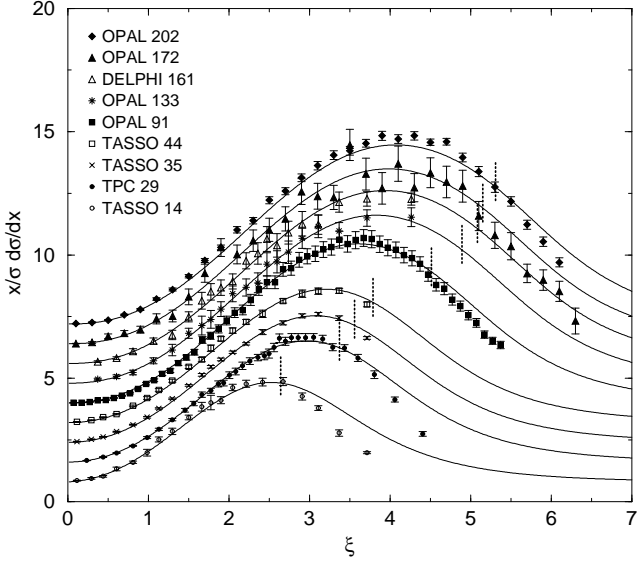}
\caption{\label{fig9} Fit to data as described in Table \ref{tab6}.}
\end{figure}

Repeating the fit with hadron mass effects accounted for (including
in the gluon jet data), we obtain $\Lambda_{\rm QCD}=490$ MeV,
$m_h=302$ MeV, and $\chi^2_{\rm DF}=2.1$. 
The fitted parameters, shown in Table \ref{tab6}, are not significantly 
different to those of Table \ref{tab5}. The comparison with the data fitted to
are shown in Figs.\ \ref{fig6} and \ref{fig9}.
The individual $\chi^2_{\rm DF}$ values
for the gluon jet data at $E_{\rm jet}=14.24$, 17.72 and 40.1 GeV
are now 1.2, 0.4 and 0.7 respectively. 
We conclude that the description of the gluon jet data 
are also improved by including hadron mass effects, particularly at low $E_{\rm jet}$.
\begin{table}[h!]
\begin{footnotesize}
\renewcommand{\arraystretch}{1.1}
\caption{\label{tab6} Parameter values for the FFs at $Q_0=14$ GeV parameterized as \
in Eq.\
(\ref{genparam}) from a fit to all data, including the OPAL 
gluon jet data
using DGLAP evolution in the DL+LO+LO$\delta$ scheme and with mass effects incorporated.
$\Lambda_{\rm QCD}=490$ MeV.}
\begin{ruledtabular}
\begin{tabular}{c|llll}
\backslashbox{FF}{Parameter} & $N$  & $\beta$ & $\alpha$ & $c$  \\
\hline
                           $g$ & 1.30 & 5.09    & $-$2.30  & 0.24 \\
\hline
                         $u+c$ & 0.46 & 1.70    & [$-$2.30]       & [0.24]   \\
\hline
                       $d+s+b$ & 0.53 & 1.75    & [$-$2.30]       & [0.24]
\end{tabular}
\end{ruledtabular}
\end{footnotesize}
\end{table}

In Fig.\ \ref{fig6}, we observe that the experimental data have a tendency to
systematically undershoot predctions at large $\xi$. This could be
partially caused by the different procedures for including soft particles in the
definitions of quark and gluon jets used in the experimental analysis. We
observe furthermore that these deviations decrease with increasing energy.

\section{Conclusions}
\label{conclusions}

We have defined a general scheme for resumming SGLs in DGLAP evolution
to any order in the FO contribution and to any class in the resummed SGL contribution. We have 
implemented it numerically at LO in the FO contribution
with DL resummation, using explicit results on the LO splitting functions from the literature. 
This DL+LO+LO$\delta$ scheme is obtained by taking the LO results for the splitting
functions, subtracting the unresummed DLs and adding the complete resummed DL contribution,
given by Eq.\ (\ref{allDLinzindelPclosed}),
which we obtained using the DLA. This scheme contains
all contributions from the MLLA, as well as all other FO contributions at LO.
We have shown by fitting to data with this scheme that for DGLAP 
evolution to also describe large $\xi$ data around the peak region, resummation of 
SGLs is necessary. In addition, we showed that both SGL resummation and treatment
of hadron mass effects leads to a good fit and a reasonable
result for $\Lambda_{\rm QCD}$.

In Ref.\ \cite{Albino:2004xa}, it was shown that the MLLA alone allows for a good description of data
around the peak region. In Ref.\ \cite{Albino:2004yg}, it was shown that provided certain
spurious higher order terms were removed, data above the peak region could also be well
described. However, the low $\xi$ region was not well reproduced, while in Ref.\ \cite{Albino:2004xa}
the good description here was due to a well known coincidence. 
Simultaneously fitting to both small and large $\xi$ data using either of the two approaches
leads to extremely high $\chi^2_{\rm DF}$ values.
Using the approach in this paper gives a much better fit to all the data, even
if the fit is still not in the acceptable range. It allows for the data in a larger 
$\xi$ range to be described
than the FO approach and the MLLA approach of Ref.\ \cite{Albino:2004xa} do. 
Further improvement in the
large $\xi$ region can be expected from the inclusion of higher order
SGLs. 

Our scheme allows a determination of quark and gluon FFs over a wider range of data
than previously achieved, and should
be incorporated in NLO global fits of FFs such as that in Ref.\ \cite{Albino:2005me} 
by using the method of the last paragraph of Sec.\ \ref{CtD},
since the current range of $0.1<x_p<1$ is very limited.

\appendix
\section{Timelike Splitting Functions in Perturbative QCD \label{app}}

The LO terms of the timelike splitting functions in $x$ space 
for the component $D=(D_{\Sigma},D_g)$ are \cite{Altarelli:1977zs}
\beq
\begin{split}
P^{(0)}_{\Sigma \Sigma}(x)=&\ C_F\left(-1-x+2\left[\frac{1}{1-x}\right]_+ + \frac{3}{2}\delta(1-x)\right),\\
P^{(0)}_{\Sigma g}=&\ 2C_F\frac{1+(1-x)^2}{x},\\
P^{(0)}_{g \Sigma}=&\ T_R n_f(x^2+(1-x)^2),\\
P^{(0)}_{gg}=&\ 2C_A \left(\frac{1}{x}-2+x-x^2+\left[\frac{1}{1-x}\right]_+\right)\\
&+\left(\frac{11}{6}C_A-\frac{2}{3}T_R n_f\right)\delta(1-x),
\end{split}
\label{A1}
\eeq
where $T_R=1/2$, $n_f$ is the number of flavours and,
for the color gauge group SU(3), $C_A=3$ and $C_F=4/3$.
The function $[1/(1-x)]_+$ is defined by
\beq
\begin{split}
&\int_x^1 \frac{dy}{y}\left[\frac{1}{1-y}\right]_+ f\left(\frac{x}{y}\right)\\
&=\int_x^1 \frac{dy}{y}\left[\frac{1}{1-y}\right]\left[f\left(\frac{x}{y}\right)-yf(x)\right]
+f(x)\ln(1-x)
\end{split}
\eeq
for any function $f(x)$. Transforming Eq.\ (\ref{A1}) to Mellin space gives
\beq
\begin{split}
P^{(0)}_{\Sigma \Sigma}(\omega)=&\ C_F
\left[\frac{3}{2}+\frac{1}{(\omega+1)(\omega+2)}-2S_1(\omega+1)\right], \\
P^{(0)}_{\Sigma g}(\omega)=&\ 2C_F
\frac{\omega^2+3\omega+4}{\omega(\omega+1)(\omega+2)}, \\
P^{(0)}_{g \Sigma}(\omega)=&\ T_R n_f
\frac{\omega^2+3\omega+4}{(\omega+1)(\omega+2)(\omega+3)}, \\
P^{(0)}_{gg}(\omega)=&\ 2C_A
\bigg[\frac{11}{12}+\frac{1}{\omega(\omega+1)}+\frac{1}{(\omega+2)(\omega+3)}\\
&-S_1(\omega+1)\bigg]-\frac{2}{3}T_R n_f,
\label{P0NLOmel}
\end{split}
\eeq
where, for integer $n$,
\beq
S_1(n)=\sum_{k=1}^n \frac{1}{k}.
\eeq

The DLs and SLs at LO are obtained by expanding 
Eq.\ (\ref{P0NLOmel}) about $\omega=0$, for which the result
\beq
S_1(\omega+1)=1+O(\omega)
\eeq
is required.

The NLO splitting functions $P^{(1)}(x)$ are presented in Ref.\ \cite{P1NLOx}, while their
Mellin transforms $P^{(1)}(\omega)$ are presented in Ref.\ \cite{Gluck:1992zx}.

We do not explicitly present $P^{(0,1)}(x)$ for the components $D=D_q^-,D_{NS}$, 
since it is enough for our purposes to know that they do not contain SGLs.

\begin{acknowledgments}

This work was supported in part by the Deutsche Forschungsgemeinschaft     
through Grant No.\ KN~365/5-1 and by the Bundesministerium f\"ur Bildung und  
Forschung through Grant No.\ 05~HT4GUA/4.

\end{acknowledgments}





\begin{thebibliography}{}

\bibitem{KKP2000}
B.~A.~Kniehl, G.~Kramer and B.~P\"otter,
Nucl.\ Phys.\ \textbf{B582} (2000) 514; 
S.~Kretzer, Phys.\ Rev.\ \textbf{D62} (2000) 054001;
L.~Bourhis, M.~Fontannaz, J.~P.~Guillet and M.~Werlen,
Eur.\ Phys.\ J.\ \textbf{C19} (2001) 89.

\bibitem{Albino:2005me}
S.~Albino, B.~A.~Kniehl and G.~Kramer,
Nucl.\ Phys.\ \textbf{B725} (2005) 181.
  
\bibitem{DGLAP}
L.~N.~Lipatov, Yad.\ Fiz.\ \textbf{20} (1974) 181 [Sov.\ J.\ Nucl.\ Phys.\ \textbf{20} (1975) 94];
V.~N.~Gribov and L.~N.~Lipatov, Yad.\ Fiz.\ \textbf{15} (1972) 781 
[Sov.\ J.\ Nucl.\ Phys.\ \textbf{15} (1972) 438];
G.\ Altarelli and G.\ Parisi, Nucl.\ Phys.\ \textbf{B126} (1977) 298;
Yu.\ L.\ Dokshitzer, Zh.\ Eksp.\ Teor.\ Fiz.\ \textbf{73} (1977) 1216
[Sov.\ Phys.\ JETP \textbf{46} (1977) 641].

\bibitem{Curci:1980uw;Furmanski:1980cm}
G.~Curci, W.~Furmanski and R.~Petronzio,
Nucl.\ Phys.\ \textbf{B175} (1980) 27;
W.~Furmanski and R.~Petronzio,
Phys.\ Lett.\ \textbf{B97} (1980) 437.

\bibitem{Bassetto:1982ma;Fadin:1983aw}
A.~Bassetto, M.~Ciafaloni, G.~Marchesini and A.~H.~Mueller,
Nucl.\ Phys.\ \textbf{B207} (1982) 189;
V.~S.~Fadin, Yad.\ Fiz.\ \textbf{37} (1983) 408 
[Sov.\ J.\ Nucl.\ Phys.\ \textbf{37} (1983) 245].

\bibitem{Dokshitzer:1991wu}
Y.~L.~Dokshitzer, V.~A.~Khoze, A.~H.~Mueller and S.~I.~Troian,
Basics of Perturbative QCD (Editions Fronti\`eres, Gif-sur-Yvette, 1991).

\bibitem{Dokshitzer:1984dx}
Y.\ L.\ Dokshitzer and S.\ I.\ Troian, Proc.\ 19th Winter School of the LNPI, Vol.\ 1,
p.\ 144 (Leningrad, 1984); Y.\ L.\ Dokshitzer and S.\ I.\ Troian, LNPI-922 preprint (1984).

\bibitem{Mueller:1982cq}
A.~H.~Mueller,
Nucl.\ Phys.\ \textbf{B213} (1983) 85.

\bibitem{Albino:2004yg}
S.~Albino, B.~A.~Kniehl and G.~Kramer,
Eur.\ Phys.\ J.\ \textbf{C38} (2004) 177.

\bibitem{Albino:2005gg}
S.~Albino, B.~A.~Kniehl, G.~Kramer and W.~Ochs,
hep-ph/0503170, DESY-05-047, MPP-2005-19,
Phys.\ Rev.\ Lett.\ ({\it in press}).

\bibitem{Collins:1998rz}
J.~C.~Collins,
Phys.\ Rev.\ \textbf{D58} (1998) 094002.

\bibitem{Albino:2004xa}
S.~Albino, B.~A.~Kniehl, G.~Kramer and W.~Ochs,
Eur.\ Phys.\ J.\ \textbf{C36} (2004) 49.

\bibitem{Braunschweig:1990yd}
W.~Braunschweig {\it et al.}  [TASSO Collaboration],
Z.\ Phys.\ \textbf{C47} (1990) 187.

\bibitem{Althoff:1983ew}
M.~Althoff {\it et al.}  [TASSO Collaboration],
Z.\ Phys.\ \textbf{C22} (1984) 307.

\bibitem{Petersen:1987bq}
A.\ Petersen {\it et al.}  [MARK II Collaboration],
Phys.\ Rev.\ \textbf{D37}, (1988) 1.

\bibitem{Aihara:1988su}
H.~Aihara {\it et al.}  [TPC/Two Gamma Collaboration],
Phys.\ Rev.\ Lett.\ \textbf{61} (1988) 1263.

\bibitem{Itoh:1994kb}
R.~Itoh {\it et al.}  [TOPAZ Collaboration],
Phys.\ Lett.\ \textbf{B345} (1995) 335.

\bibitem{Barate:1996fi}
ALEPH, R.\ Barate {\it et al.}  [ALEPH Collaboration],
Phys.\ Rept.\ \textbf{294} (1998) 1.

\bibitem{Abreu:1996na}
DELPHI, P.\ Abreu {\it et al.}  [DELPHI Collaboration],
Z.\ Phys.\ \textbf{C73} (1996) 11.

\bibitem{Adeva:1991it}
B.\ Adeva {\it et al.}  [L3 Collaboration],
Phys.\ Lett.\ \textbf{B259} (1991) 199.

\bibitem{Akrawy:1990ha}
M.~Z.~Akrawy {\it et al.}  [OPAL Collaboration],
Phys.\ Lett.\ \textbf{B247} (1990) 617.

\bibitem{Abrams:1989rz}
G.~S.~Abrams {\it et al.}  [SLC Collaboration],
Phys.\ Rev.\ Lett.\  \textbf{64} (1990) 1334.

\bibitem{Buskulic:1996tt}
D.\ Buskulic {\it et al.}  [ALEPH Collaboration],
Z.\ Phys.\ \textbf{C73} (1997) 409.

\bibitem{Alexander:1996kh}
G.~Alexander {\it et al.}  [OPAL Collaboration],
Z.\ Phys.\ \textbf{C72} (1996) 191.

\bibitem{Ackerstaff:1997kk}
K.\ Ackerstaff {\it et al.}  [OPAL Collaboration],
Z.\ Phys.\ \textbf{C75}, (1997) 193.

\bibitem{Abbiendi:1999sx}
G.~Abbiendi {\it et al.}  [OPAL Collaboration],
Eur.\ Phys.\ J.\ \textbf{C16} (2000) 185.

\bibitem{Abbiendi:2002mj}
G.~Abbiendi {\it et al.}  [OPAL Collaboration],
Eur.\ Phys.\ J.\ \textbf{C27} (2003) 467.

\bibitem{Lupia:1997hj}
S.~Lupia and W.~Ochs,
Eur.\ Phys.\ J.\ \textbf{C2} (1998) 307.

\bibitem{Azimov:1985by}
Y.~I.~Azimov, Y.~L.~Dokshitzer, V.~A.~Khoze and S.~I.~Troian,
Z.\ Phys.\ \textbf{C31} (1986) 213.

\bibitem{Abbiendi:2003gh}
G.~Abbiendi {\it et al.}  [OPAL Collaboration],
Phys.\ Rev.\ \textbf{D69} (2004) 032002.

\bibitem{Abbiendi:1999pi}
G.~Abbiendi {\it et al.}  [OPAL Collaboration],
Eur.\ Phys.\ J.\ \textbf{C11} (1999) 217.

\bibitem{Altarelli:1977zs}
G.~Altarelli and G.~Parisi,
Nucl.\ Phys.\ \textbf{B126} (1977) 298.

\bibitem{P1NLOx}
G.~Curci, W.~Furmanski and R.~Petronzio,
Nucl.\ Phys.\ \textbf{B175} (1980) 27;
W.~Furmanski and R.~Petronzio,
Phys.\ Lett.\ \textbf{B97} (1980) 437;
E.~G.~Floratos, C.~Kounnas and R.~Lacaze,
Nucl.\ Phys.\ \textbf{B192} (1981) 417;
J.~Binnewies, B.~A.~Kniehl and G.~Kramer,
Z.\ Phys.\ \textbf{C76} (1997) 677;
R.~K.~Ellis, W.~J.~Stirling and B.~R.~Webber,
``QCD and collider physics,''
Camb.\ Monogr.\ Part.\ Phys.\ Nucl.\ Phys.\ Cosmol.\  \textbf{8} (1996) 1, Chapter 6.
In the latter reference, some misprints from the earlier publications are corrected.

\bibitem{Gluck:1992zx}
M.~Gluck, E.~Reya and A.~Vogt,
Phys.\ Rev.\ \textbf{D48} (1993) 116
[Erratum-ibid.\ \textbf{D51} (1995) 1427].

\end{thebibliography}
\end{document}